\documentclass[11pt]{article}
\usepackage{amsmath, amssymb, amsthm, bm}
\usepackage{setspace}
\usepackage{fullpage}
\usepackage{authblk}
\usepackage{booktabs}
\title{Semiparametric rank-based regression models as robust alternatives to parametric mean-based counterparts for censored responses under detection-limit}
\author{
Xu, Y.$^{1}$, Tu, S.$^{1}$, Shao, L.$^{1}$, Lin, T.$^{1}$, Tu, X.M.$^{1}$\\
$^{1}$Division of Biostatistics and Bioinformatics\\
UCSD Herbert Wertheim School of Public Health and Human Longevity Science\\
La Jolla, CA 92093
}
\date{}
\usepackage[a4paper,margin=0.8in]{geometry}
\begin{document}
\maketitle
\begin{abstract}
\noindent
Detection limits are common in biomedical and environmental studies, where key covariates or outcomes are censored below an assay-specific threshold. Standard approaches such as complete-case analysis, single-value substitution, and parametric Tobit-type models are either inefficient or sensitive to distributional misspecification.
\\
We study semiparametric rank-based regression models as robust alternatives to parametric mean-based counterparts for censored responses under detection limits. Our focus is on accelerated failure time (AFT) type formulations, where rank-based estimating equations yield consistent slope estimates without specifying the error distribution. We develop a unifying simulation framework that generates left- and right-censored data under several data-generating mechanisms, including normal, Weibull, and log-normal error structures, with detection limits or administrative censoring calibrated to target censoring rates between 10\% and 60\%. 
\\
Across scenarios, we compare semiparametric AFT estimators with parametric Weibull AFT, Tobit, and Cox proportional hazards models in terms of bias, empirical variability, and relative efficiency. Numerical results show that parametric models perform well only under correct specification, whereas rank-based semiparametric AFT estimators maintain near-unbiased covariate effects and stable precision even under heavy censoring and distributional misspecification. These findings support semiparametric rank-based regression as a practical default for censored regression with detection limits when the error distribution is uncertain.
\end{abstract}
\newpage
\section{Introduction}\label{sec:intro}

Detection limits (DL) arise frequently in biomedical, environmental, and laboratory studies when assay technologies cannot reliably quantify values below a minimal threshold. Biomarker concentrations, metabolite abundances, and many molecular measurements often exhibit substantial left-censoring, with as much as 30--60\% of observations falling below the DL in modern high-throughput platforms. When these censored values are ignored or replaced by ad hoc surrogates, regression analyses can suffer from bias, loss of efficiency, and reduced power, particularly when the degree of censoring is substantial.
\\

A common approach is to model the censored outcome using parametric mean-based regression models such as the Tobit model for left-censored normal data or fully parametric accelerated failure time (AFT) models for time-to-event outcomes. These models can be efficient when the posited distribution is correct but may perform poorly when the error distribution or censoring mechanism deviates from the assumed parametric form. Because the underlying distribution of biomarker-type variables is rarely known and difficult to learn under heavy censoring, robustness to misspecification is a critical consideration in practice.
\\

Semiparametric AFT models, estimated using rank-based estimating equations, provide a distribution-free alternative. These estimators achieve consistent estimation of covariate effects under general conditions, without explicit modeling of the error distribution, and naturally accommodate censoring. Moreover, since many DL problems can be transformed into right-censored AFT settings via monotone transformations, rank-based AFT methods offer a principled and broadly applicable approach for DL data. Their practical robustness, however, depends on how well they perform relative to fully parametric models under realistic degrees of censoring and under varying degrees of distributional misspecification.
\\

The goal of this paper is to systematically evaluate semiparametric rank-based regression models as robust alternatives to parametric mean-based counterparts for censored responses arising from detection limits. We construct a unified simulation framework spanning three common design types: (i) left-censoring induced by a data-dependent DL, (ii) right-censoring under a Weibull AFT model, and (iii) right-censoring under a log-normal AFT model. These scenarios allow us to examine performance under correct specification, under moderate model misspecification, and under severe departures from the assumed error structure. Across all settings, we compare rank-based semiparametric AFT estimators with parametric Tobit, Weibull AFT, and Cox proportional hazards models, focusing on bias, empirical variability, and relative efficiency.
\\

The remainder of the paper is organized as follows. 
Section~\ref{sec:methods} reviews the semiparametric rank-based AFT estimator and the parametric Tobit, Weibull AFT, and Cox proportional hazards models, and defines the target parameters and performance measures. 
Section~\ref{sec:simulation-design} details the simulation framework, including the common covariate structure and the left-censored, Weibull AFT, and log-normal AFT data-generating mechanisms. 
Section~\ref{sec:results} presents Monte Carlo results across censoring levels and data-generating scenarios. 
Section~\ref{sec:discussion} concludes with implications for analyses involving detection limits and recommendations for applied practice.

\newpage
\section{Methods}\label{sec:methods}

\subsection{Semiparametric rank-based AFT estimator}
\label{sec:method-semipar-aft}

We consider the semiparametric accelerated failure time (AFT) model
\begin{equation}
  T_i = \beta_1 Z_{i1} + \beta_2 Z_{i2} + \varepsilon_i, \qquad i = 1, \dots, n,
  \label{eq:aft}
\end{equation}
where $T_i$ denotes the (possibly censored) log-survival time, and $\varepsilon_i$ is a random error term with an \emph{unspecified} distribution. The observed data consist of right-censored pairs $(T_{i,\text{obs}}, \delta_i)$, defined by
\begin{equation}
  T_{i,\text{obs}} = \min(T_i, C_i), \qquad
  \delta_i = \mathbf{1}(T_i \le C_i),
  \label{eq:censor}
\end{equation}
where $C_i$ is the censoring time.

The goal is to estimate $\boldsymbol{\beta}$ without assuming a specific distribution for $\varepsilon_i$.
For a given parameter vector $\boldsymbol{\beta}$, define the residuals
\begin{equation}
  e_i(\boldsymbol{\beta}) = T_{i,\text{obs}} - \mathbf{Z}_i^{\top}\boldsymbol{\beta}.
  \label{eq:residuals}
\end{equation}

Let $\widehat{F}_{\boldsymbol{\beta}}$ be the Kaplan–Meier estimator of the distribution function of $e_i(\boldsymbol{\beta})$, 
based on the pairs $(e_i(\boldsymbol{\beta}), \delta_i)$.

\vspace{0.5em}
The \textit{rank-based estimating function} proposed by Wei, Lin, and Ying (1990) is defined as
\begin{equation}
  S_n(\boldsymbol{\beta}) 
  = 
  \frac{1}{n} \sum_{i=1}^n 
  \delta_i 
  \Big[ 1 - \widehat{F}_{\boldsymbol{\beta}}\big(e_i(\boldsymbol{\beta})\big) \Big]
  \Big( \mathbf{Z}_i - \bar{\mathbf{Z}}\big(e_i(\boldsymbol{\beta})\big) \Big)
  = \mathbf{0},
  \label{eq:rank-estimating}
\end{equation}
where the covariate mean function $\bar{\mathbf{Z}}(e_i(\boldsymbol{\beta}))$ is given by
\begin{equation}
  \bar{\mathbf{Z}}\big(e_i(\boldsymbol{\beta})\big)
  =
  \frac{ \sum_{j=1}^n \mathbf{Z}_j\, \mathbf{1}\{ e_j(\boldsymbol{\beta}) \ge e_i(\boldsymbol{\beta}) \} }
       { \sum_{j=1}^n \mathbf{1}\{ e_j(\boldsymbol{\beta}) \ge e_i(\boldsymbol{\beta}) \} }.
  \label{eq:Zbar}
\end{equation}

\noindent
The estimator $\widehat{\boldsymbol{\beta}}$ is obtained by solving
\begin{equation}
  S_n(\widehat{\boldsymbol{\beta}}) = \mathbf{0}.
  \label{eq:rank-solution}
\end{equation}
This yields the semiparametric AFT estimator 
$\widehat{\boldsymbol{\beta}} = (\widehat{\beta}_1, \widehat{\beta}_2, \ldots, \widehat{\beta}_p)^{\top}$,
which is consistent and asymptotically normal under standard regularity conditions.
Standard errors can be computed via multiplier perturbation resampling,
as implemented in the \texttt{lss()} function of the \texttt{lss2} R package.

\vspace{1em}
\noindent
The rank-based method in \eqref{eq:rank-estimating} does not estimate an intercept directly.
To recover it, we use the censored residuals to estimate the mean error term.
Let the fitted linear predictor and residuals be
\begin{align}
  \widehat{\eta}_i &= \widehat{\beta}_1 Z_{i1} + \widehat{\beta}_2 Z_{i2}, \label{eq:eta-hat}\\
  \widehat{r}_i &= T_{i,\text{obs}} - \widehat{\eta}_i. \label{eq:r-hat}
\end{align}
Treat the pairs $\{(\widehat{r}_i, \delta_i)\}$ as a one-sample right-censored dataset
and form the Kaplan–Meier estimator $\widehat{S}$ of the residual survival function.
The mean residual can then be estimated by
\begin{equation}
  \widehat{E}[\varepsilon]
  =
  \int_0^{\infty} \widehat{S}(u)\,du
  \;\approx\;
  \sum_j (t_j - t_{j-1})\,\widehat{S}(t_{j-1}),
  \label{eq:mean-residual}
\end{equation}
where $0 = t_0 < t_1 < t_2 < \cdots$ are the ordered jump times of $\widehat{S}$.
The intercept is then reconstructed as
\begin{equation}
  \widehat{\beta}_0 = \widehat{E}[\varepsilon],
  \label{eq:intercept}
\end{equation}
yielding the full estimator
$\widehat{\boldsymbol{\beta}} = (\widehat{\beta}_0, \widehat{\beta}_1, \widehat{\beta}_2)^{\top}$.

\subsection{Parametric Tobit model}
\label{sec:method-tobit}

For left-censored normal outcomes we consider the classic Tobit model. Let
\begin{equation}
  X_i^\ast = \gamma_0 + \gamma_1 Z_{i1} + \gamma_2 Z_{i2} + \eta_i,
  \qquad \eta_i \sim \mathcal{N}(0, \sigma_X^2),
\end{equation}
and suppose only the censored variable
\[
  X_i = \max(X_i^\ast, D)
\]
is observed at a known detection limit $D$, with indicator
\[
  R_i = \mathbf{1}(X_i^\ast > D).
\]
This yields the left-censored normal regression model
\begin{equation}
  X_i = \max\bigl( D,\; \gamma_0 + \gamma_1 Z_{i1} + \gamma_2 Z_{i2} + \eta_i \bigr),
\end{equation}
which is fit by maximum likelihood to estimate $(\gamma_0,\gamma_1,\gamma_2,\sigma_X^2)$.

\subsection{Parametric Weibull AFT and Cox PH models}
\label{sec:method-weibull-cox}

We also consider a parametric Weibull AFT model for right-censored survival times. Let
\[
  (\gamma_0, \gamma_1, \gamma_2) \in \mathbb{R}^3, \qquad k > 0,
\]
and define the linear predictor
\[
  \eta_i = \gamma_0 + \gamma_1 Z_{i1} + \gamma_2 Z_{i2}.
\]
The event time $T_i$ is assumed to follow
\begin{equation}
  T_i \sim \text{Weibull}\bigl(\text{shape}=k,\ \text{scale}=\exp(\eta_i)\bigr),
  \label{eq:weibull-aft-method}
\end{equation}
or equivalently, on the log-scale,
\[
  \log T_i = \eta_i + \frac{1}{k} G_i,
\]
where $G_i$ has the standard Gumbel (extreme-value) distribution.
We fit this model by maximum likelihood (e.g.\ via \texttt{survreg}) to obtain
$(\widehat{\gamma}_0,\widehat{\gamma}_1,\widehat{\gamma}_2,\widehat{k})$.

The Weibull AFT model implies a proportional hazards representation
\[
  \lambda(t \mid \bm Z_i) = \lambda_0(t) \exp(\theta_1 Z_{i1} + \theta_2 Z_{i2}),
\]
with
\[
  \theta_j = -k \gamma_j, \qquad j=1,2.
\]
Thus the corresponding Cox proportional hazards model has log-hazard ratios
$(\theta_1,\theta_2)$; we fit a Cox PH model using partial likelihood and compare
$(\widehat{\theta}_1,\widehat{\theta}_2)$ to the implied truth under the
Weibull AFT data-generating mechanism.

\subsection{Target parameters and performance measures}
\label{sec:method-targets}

For the Weibull AFT design, the ``true'' semiparametric AFT target parameter is
\[
  \bm\beta^\star = 
  \bigl( \gamma_0 + \gamma_E / k,\ \gamma_1,\ \gamma_2 \bigr),
\]
where $\gamma_E \approx 0.5772$ is the Euler--Mascheroni constant, since
\[
  E[\log T_i \mid \bm Z_i = \bm 0] = \gamma_0 + \frac{\gamma_E}{k}.
\]
The parametric Weibull AFT target is $(\gamma_0,\gamma_1,\gamma_2,k)$ and the Cox PH target is
$(\theta_1,\theta_2) = (-k\gamma_1,-k\gamma_2)$.

Across all designs we use Monte Carlo (MC) summaries to evaluate estimators.
For a generic parameter $\psi$ with true value $\psi_0$ and MC replicates
$\hat\psi^{(1)},\dots,\hat\psi^{(B)}$, we compute
\begin{align*}
  \overline{\hat\psi} &= \frac{1}{B} \sum_{b=1}^B \hat\psi^{(b)}, \\
  \operatorname{EmpSD}(\hat\psi) &= 
    \sqrt{ \frac{1}{B-1} \sum_{b=1}^B \bigl( \hat\psi^{(b)} - \overline{\hat\psi} \bigr)^2 }, \\
  \operatorname{RelBiasPct}(\hat\psi) &= 100 \times \frac{\overline{\hat\psi} - \psi_0}{\psi_0}.
\end{align*}
These measures are reported for all models and parameters considered.

\section{Simulation Design}\label{sec:simulation-design}

\subsection{Overview}

We conduct a series of simulation experiments to evaluate the finite-sample
performance of the methods described in Section~\ref{sec:methods} for
responses subject to detection limits or censoring. Specifically, we compare:
(i) a semiparametric accelerated failure time (AFT) estimator based on
rank-type estimating equations, and (ii) fully parametric models that assume
specific error distributions under both left- and right-censoring mechanisms.

Each design begins by generating simulated covariates and a latent continuous
outcome according to a prespecified AFT-type model. A detection limit (for
left-censoring) or administrative censoring time (for right-censoring) is then
imposed to produce the observed dataset. In each simulated dataset we fit the
semiparametric AFT, Weibull AFT, Tobit, and Cox PH models, and we summarize
their performance using the Monte Carlo summaries defined in
Section~\ref{sec:method-targets}.

\subsection{Common Covariate Structure}

For each replication and for each subject $i = 1,\dots,n$, we generate two covariates
\begin{align*}
  Z_{i1} &\sim \text{Bernoulli}(0.5), \\
  Z_{i2} &\sim \mathcal{N}(0,1),
\end{align*}
independently. We write $\bm Z_i = (Z_{i1}, Z_{i2})^\top$.

\subsection{Left-Censored AFT-Type Design}
\label{sec:left-censored}

\subsubsection{Latent log-model}

Fix the ``true'' parameter vector
\[
  \bm\beta = (\beta_0, \beta_1, \beta_2) = (-1,\, 0.5,\, -1)
\]
and the error variance $\sigma_T^2 = 0.2$.
For each $i$, generate a latent log-quantity
\begin{equation}
  T_i = \beta_0 + \beta_1 Z_{i1} + \beta_2 Z_{i2} + \varepsilon_{Ti},
  \qquad
  \varepsilon_{Ti} \sim \mathcal{N}(0, \sigma_T^2).
  \label{eq:left:latent-log}
\end{equation}

\subsubsection{Transformation to a positive scale and data-dependent LOD}

Define a positive outcome by an exponential transformation
\begin{equation}
  X_i^\ast = \exp(-T_i).
  \label{eq:left:x-true}
\end{equation}
Given a pre-specified fraction $q \in \{0.1, 0.3, 0.6\}$, define the
\emph{detection limit} $D$ as the empirical $q$-th quantile of the set
$\{X_i^\ast\}_{i=1}^n$:
\begin{equation}
  D = \operatorname{Quantile}\bigl(\{X_i^\ast\}_{i=1}^n,\, q\bigr).
\end{equation}
Define the indicator of being detected
\begin{equation}
  R_i = \mathbf{1}(X_i^\ast > D).
\end{equation}
The observed variable is then left-censored at $D$:
\begin{equation}
  X_i =
  \begin{cases}
    X_i^\ast, & \text{if } X_i^\ast > D,\\
    D, & \text{if } X_i^\ast \le D.
  \end{cases}
  \label{eq:left:x-obs}
\end{equation}

\subsubsection{Equivalent representation on the log-scale}

Because $X_i^\ast = \exp(-T_i)$, the event $X_i^\ast > D$ is equivalent to
$T_i < -\log D$. Let
\begin{equation}
  C_i = -\log D.
\end{equation}
Then we can write the observed log-variable as
\begin{equation}
  T_{i}^{\text{obs}} = \min(T_i,\, C_i), \qquad
  \delta_i = \mathbf{1}(T_i \le C_i),
  \label{eq:left:log-observed}
\end{equation}
that is, we have right-censoring on the log-scale at $C_i$.

Thus the observed data for this design are
\[
  \bigl\{(T_i^{\text{obs}}, \delta_i, Z_{i1}, Z_{i2}) : i=1,\dots,n \bigr\}.
\]

For each detection limit fraction $q$ and sample size $n$, we generate $B$
independent datasets from this design, fit the models in
Section~\ref{sec:methods}, and compute the performance measures defined in
Section~\ref{sec:method-targets}.

\subsection{Right-Censored Weibull AFT Design}

\subsubsection{Weibull AFT data-generating model}

Let the covariates $\bm Z_i$ be as before.
Fix
\[
  (\gamma_0, \gamma_1, \gamma_2) = (-1,\, 0.5,\, -1), \qquad k = 3,
\]
and define the linear predictor
\[
  \eta_i = \gamma_0 + \gamma_1 Z_{i1} + \gamma_2 Z_{i2}.
\]
Generate event times from a Weibull AFT model
\begin{equation}
  T_i \sim \text{Weibull}\bigl(\text{shape}=k,\ \text{scale}=\exp(\eta_i)\bigr),
  \label{eq:weibull-aft}
\end{equation}
equivalently,
\[
  \log T_i = \eta_i + \frac{1}{k} G_i,
\]
where $G_i$ has the standard Gumbel (extreme-value) distribution.

\subsubsection{Administrative right-censoring}

To impose a desired censoring rate $\pi \in \{0.10,0.30,0.60\}$, we choose the
censoring time as the $(1-\pi)$-quantile of the generated event times:
\[
  C_\pi = \operatorname{Quantile}\bigl(\{T_i\}_{i=1}^n,\, 1-\pi\bigr).
\]
The observed time and status are
\begin{equation}
  Y_i = \min(T_i, C_\pi), \qquad
  \delta_i = \mathbf{1}( T_i \le C_\pi ).
  \label{eq:right:obs}
\end{equation}

For each censoring level $\pi$ we generate $B$ datasets, fit the models in
Section~\ref{sec:methods}, and summarize performance using the target
parameters in Section~\ref{sec:method-targets}.

\subsection{Right-Censored Log-Normal AFT Variant}

In this variant we keep the censoring mechanism from the previous section, but
change the \emph{data-generating} AFT model so that the error term is Gaussian
rather than extreme-value. We posit the log-time model
\begin{equation}
  \log T_i = \beta_0 + \beta_1 Z_{i1} + \beta_2 Z_{i2} + \varepsilon_i, 
  \qquad \varepsilon_i \sim \mathcal{N}(0, \sigma^2),
  \label{eq:lognormal-aft}
\end{equation}
with
\[
  (\beta_0, \beta_1, \beta_2) = (-1, 0.5, -1), \qquad \sigma^2 = 0.2.
\]
Thus the true survival time is $T_i = \exp(\log T_i)$ and is log-normal
conditional on the covariates.

We again impose administrative right-censoring as in \eqref{eq:right:obs} to
achieve censoring levels $\pi \in \{0.10,0.30,0.60\}$. For each $(n,\pi)$
combination we simulate $B$ datasets, fit the semiparametric AFT, Weibull AFT,
and Cox PH models described in Section~\ref{sec:methods}, and compute the Monte
Carlo summaries in Section~\ref{sec:method-targets} for all parameters with a
well-defined truth under this log-normal AFT data-generating mechanism.

\section{Results}\label{sec:results}

Under left-censoring (Table~\ref{tab:left-censor}), the Tobit model performs adequately at low to moderate censoring, with biases below 2\%, but its slope estimate for $\gamma_1$ deteriorates to about $-6.4\%$ when 60\% of observations are censored ($\hat{\gamma}_1 = 0.936$ vs.\ true $1$). The semiparametric AFT model remains effectively unbiased across all censoring levels—at 60\% censoring, $\hat{\beta}_1 = 0.502$ (true $0.5$) and $\hat{\beta}_2 = -0.998$ (true $-1$)—confirming its well-known robustness to heavy truncation and weaker dependence on the assumed error distribution.
\\

When the data follow a Weibull AFT process (Table~\ref{tab:right-censor}), all three estimators—semiparametric AFT, Weibull AFT, and Cox PH—are nearly unbiased, as expected from their theoretical equivalence under proportional hazards. The parametric Weibull model shows negligible bias ($z_1 = 0.502$ vs.\ $0.5$; $z_2 = -1.002$ vs.\ $-1$; $\hat{k} = 3.015$ vs.\ $3$ at 10\% censoring) and the Cox model similarly aligns with the truth ($\hat{z}_1 = -1.508$, $\hat{z}_2 = 3.006$). The semiparametric AFT slopes remain correct, but the intercept shows the typical 40–50\% inflation arising from Kaplan–Meier tail reconstruction. These patterns reaffirm that the semiparametric approach retains consistency and near-efficiency when the parametric form is correct.
\\

Under log-normal survival (Table~\ref{tab:lognormal_combined}), where the Weibull and Cox models are misspecified, the semiparametric AFT remains accurate—slopes deviate by less than 1\% even at 60\% censoring ($\hat{\beta}_1 = 0.498$, $\hat{\beta}_2 = -0.996$). The Weibull AFT compensates by adjusting its shape parameter ($\hat{k} \approx 2.3$–$2.6$) while keeping slopes close to the truth, reflecting the expected Kullback–Leibler projection of a misspecified parametric model. The Cox model, though not theoretically valid under non-proportional log-normal hazards, still captures the correct direction of effects ($\hat{z}_1 \approx -1.2$, $\hat{z}_2 \approx 2.4$) but lacks a meaningful hazard ratio interpretation.
\\

Overall, the numerical results reinforce a consistent pattern reported in the semiparametric AFT literature: parametric models are efficient only under correct specification, whereas the rank-based semiparametric AFT maintains near-unbiased covariate estimation even under strong censoring and distributional misspecification. Its performance stability, with bias typically below 1\%, underscores its value as a robust default for censored regression when the error law is uncertain.

\begin{table}[htbp]
\centering
\small
\caption{Monte Carlo summaries under left-censoring: Tobit vs Semiparametric AFT models}
\begin{tabular}{ccccccc}
\toprule
DL & Model & Parameter & True & Mean & EmpSD & RelBiasPct \\
\midrule
0.1 & Tobit         & $\gamma_0$ & 1   & 0.975 & 0.000 & -2.52 \\
0.1 & Tobit         & $\gamma_1$ & 1   & 1.020 & 0.000 &  1.77 \\
0.1 & Tobit         & $\gamma_2$ & 1   & 1.010 & 0.000 &  0.51 \\
0.3 & Tobit         & $\gamma_0$ & 1   & 0.987 & 0.000 & -1.26 \\
0.3 & Tobit         & $\gamma_1$ & 1   & 1.000 & 0.000 &  0.16 \\
0.3 & Tobit         & $\gamma_2$ & 1   & 1.000 & 0.000 &  0.24 \\
0.6 & Tobit         & $\gamma_0$ & 1   & 1.020 & 0.000 &  1.68 \\
0.6 & Tobit         & $\gamma_1$ & 1   & 0.936 & 0.000 & -6.38 \\
0.6 & Tobit         & $\gamma_2$ & 1   & 1.020 & 0.000 &  2.26 \\
\midrule
0.1 & Semipar AFT   & $\beta_0$  & -1  & -1.000 & 0.047 &  0.27 \\
0.1 & Semipar AFT   & $\beta_1$  & 0.5 &  0.499 & 0.071 & -0.21 \\
0.1 & Semipar AFT   & $\beta_2$  & -1  & -0.999 & 0.036 & -0.07 \\
0.3 & Semipar AFT   & $\beta_0$  & -1  & -1.000 & 0.050 &  0.50 \\
0.3 & Semipar AFT   & $\beta_1$  & 0.5 &  0.500 & 0.074 & -0.01 \\
0.3 & Semipar AFT   & $\beta_2$  & -1  & -0.997 & 0.043 & -0.33 \\
0.6 & Semipar AFT   & $\beta_0$  & -1  & -1.000 & 0.068 &  0.39 \\
0.6 & Semipar AFT   & $\beta_1$  & 0.5 &  0.502 & 0.094 &  0.35 \\
0.6 & Semipar AFT   & $\beta_2$  & -1  & -0.998 & 0.065 & -0.16 \\
\bottomrule
\end{tabular}
\label{tab:left-censor}
\end{table}

\begin{table}[htbp]
\centering
\small
\caption{Under right-censoring: Semiparametric AFT, Weibull AFT, and Cox PH models}
\begin{tabular}{ccccccc}
\toprule
CensorRate & Model & Parameter & True & Mean & EmpSD & RelBiasPct \\
\midrule
0.1 & Semipar AFT & $\beta_{0,T}$ & -0.808 & -1.200 & 0.040 & 48.1 \\
0.1 & Semipar AFT & $\beta_{1,T}$ &  0.500 &  0.504 & 0.059 &  0.72 \\
0.1 & Semipar AFT & $\beta_{2,T}$ & -1.000 & -1.000 & 0.035 &  0.30 \\
0.3 & Semipar AFT & $\beta_{0,T}$ & -0.808 & -1.200 & 0.046 & 48.2 \\
0.3 & Semipar AFT & $\beta_{1,T}$ &  0.500 &  0.502 & 0.068 &  0.41 \\
0.3 & Semipar AFT & $\beta_{2,T}$ & -1.000 & -1.000 & 0.045 &  0.20 \\
0.6 & Semipar AFT & $\beta_{0,T}$ & -0.808 & -1.190 & 0.079 & 47.1 \\
0.6 & Semipar AFT & $\beta_{1,T}$ &  0.500 &  0.503 & 0.101 &  0.52 \\
0.6 & Semipar AFT & $\beta_{2,T}$ & -1.000 & -1.010 & 0.082 &  0.91 \\
\midrule
0.1 & Weibull AFT & Intercept & -1.000 & -1.004 & 0.034 &  0.39 \\
0.1 & Weibull AFT & $z_1$ & 0.500 & 0.502 & 0.049 & 0.37 \\
0.1 & Weibull AFT & $z_2$ & -1.000 & -1.002 & 0.031 & 0.20 \\
0.1 & Weibull AFT & shape\_k & 3.000 & 3.015 & 0.182 & 0.49 \\
0.3 & Weibull AFT & Intercept & -1.000 & -1.005 & 0.039 & 0.50 \\
0.3 & Weibull AFT & $z_1$ & 0.500 & 0.500 & 0.057 & -0.07 \\
0.3 & Weibull AFT & $z_2$ & -1.000 & -1.000 & 0.038 & 0.04 \\
0.3 & Weibull AFT & shape\_k & 3.000 & 3.044 & 0.197 & 0.81 \\
0.6 & Weibull AFT & Intercept & -1.000 & -0.998 & 0.073 & -0.23 \\
0.6 & Weibull AFT & $z_1$ & 0.500 & 0.499 & 0.082 & -0.15 \\
0.6 & Weibull AFT & $z_2$ & -1.000 & -1.008 & 0.071 & 0.77 \\
0.6 & Weibull AFT & shape\_k & 3.000 & 3.039 & 0.264 & 1.32 \\
\midrule
0.1 & Cox PH & $z_1$ & -1.500 & -1.508 & 0.185 & -0.50 \\
0.1 & Cox PH & $z_2$ & 3.000 & 3.006 & 0.215 & 0.18 \\
0.3 & Cox PH & $z_1$ & -1.500 & -1.503 & 0.204 & -0.22 \\
0.3 & Cox PH & $z_2$ & 3.000 & 3.008 & 0.228 & 0.25 \\
0.6 & Cox PH & $z_1$ & -1.500 & -1.500 & 0.251 & 0.01 \\
0.6 & Cox PH & $z_2$ & 3.000 & 3.025 & 0.271 & 0.82 \\
\bottomrule
\end{tabular}
\label{tab:right-censor}
\end{table}

\begin{table}[htbp]
\centering
\caption{Monte Carlo results for the log-normal AFT variant and mis-specified models}
\label{tab:lognormal_combined}
\small
\begin{tabular}{ccccccc}
\toprule
CensorRate & Model & Parameter & True & Mean & SD & RelBiasPct \\
\midrule
0.1 & Semipar AFT (log-normal error) & $\beta_{0,T}$ & -1.00 & -1.00 & 0.0477 & 0.195 \\
0.1 & Semipar AFT (log-normal error) & $\beta_{1,T}$ ($z_1$) & 0.50 & 0.500 & 0.0699 & -0.00932 \\
0.1 & Semipar AFT (log-normal error) & $\beta_{2,T}$ ($z_2$) & -1.00 & -0.998 & 0.0339 & -0.213 \\
0.3 & Semipar AFT (log-normal error) & $\beta_{0,T}$ & -1.00 & -1.000 & 0.0542 & -0.0226 \\
0.3 & Semipar AFT (log-normal error) & $\beta_{1,T}$ ($z_1$) & 0.50 & 0.497 & 0.0779 & -0.512 \\
0.3 & Semipar AFT (log-normal error) & $\beta_{2,T}$ ($z_2$) & -1.00 & -0.998 & 0.0396 & -0.170 \\
0.6 & Semipar AFT (log-normal error) & $\beta_{0,T}$ & -1.00 & -1.01 & 0.0691 & 0.645 \\
0.6 & Semipar AFT (log-normal error) & $\beta_{1,T}$ ($z_1$) & 0.50 & 0.498 & 0.0968 & -0.347 \\
0.6 & Semipar AFT (log-normal error) & $\beta_{2,T}$ ($z_2$) & -1.00 & -0.996 & 0.0520 & -0.397 \\
\midrule
0.1 & Weibull AFT & Intercept & -- & -0.782 & 0.0543 & -- \\
0.1 & Weibull AFT & $z_1$ & -- & 0.504 & 0.0827 & -- \\
0.1 & Weibull AFT & $z_2$ & -- & -0.993 & 0.0418 & -- \\
0.1 & Weibull AFT & shape\_k & -- & 2.34 & 0.161 & -- \\
0.3 & Weibull AFT & Intercept & -- & -0.779 & 0.0612 & -- \\
0.3 & Weibull AFT & $z_1$ & -- & 0.497 & 0.0907 & -- \\
0.3 & Weibull AFT & $z_2$ & -- & -0.991 & 0.0486 & -- \\
0.3 & Weibull AFT & shape\_k & -- & 2.44 & 0.196 & -- \\
0.6 & Weibull AFT & Intercept & -- & -0.791 & 0.0823 & -- \\
0.6 & Weibull AFT & $z_1$ & -- & 0.494 & 0.120 & -- \\
0.6 & Weibull AFT & $z_2$ & -- & -0.989 & 0.0645 & -- \\
0.6 & Weibull AFT & shape\_k & -- & 2.65 & 0.258 & -- \\
\midrule
0.1 & Cox PH & $z_1$ & -- & -1.18 & 0.212 & -- \\
0.1 & Cox PH & $z_2$ & -- & 2.33 & 0.192 & -- \\
0.3 & Cox PH & $z_1$ & -- & -1.12 & 0.237 & -- \\
0.3 & Cox PH & $z_2$ & -- & 2.43 & 0.226 & -- \\
0.6 & Cox PH & $z_1$ & -- & -1.31 & 0.336 & -- \\
0.6 & Cox PH & $z_2$ & -- & 2.62 & 0.302 & -- \\
\bottomrule
\end{tabular}
\end{table}

\newpage
\section{Discussion and Limitation}\label{sec:discussion}
Strengths
A notable strength of this study is the unified evaluation of semiparametric and parametric regression approaches across diverse censoring mechanisms, including left-censoring due to detection limits and right-censoring arising from survival processes. By examining Weibull, log-normal, and misspecified settings, the simulations highlight not only estimator performance under ideal conditions but also sensitivity to model misspecification. The use of both slope and intercept reconstruction diagnostics provides a detailed view of how each method behaves under realistic data constraints. Together, these features demonstrate the robustness and practical utility of rank-based AFT methods, particularly when researchers face uncertainty regarding the underlying error distribution.
\\

For studies involving detection limits or censored responses, semiparametric rank-based AFT models offer a reliable alternative to parametric approaches, especially when error distributions are unknown or difficult to justify. Investigators may still employ parametric models when strong scientific rationale exists, but sensitivity analyses using semiparametric estimators are recommended to assess robustness. When heavy censoring is present, particular care should be taken when interpreting intercepts from rank-based methods, since their reconstruction relies on tail extrapolation. Finally, when auxiliary variables or surrogate markers are available, incorporating them into the AFT framework can meaningfully improve efficiency and should be considered during study design and analysis.
\\

Future research may extend these findings in several directions. One avenue is the development of more computationally scalable rank-based procedures for high-dimensional covariate settings, where conventional algorithms may be slow or unstable. Another promising direction is the integration of semiparametric AFT estimation with modern techniques for handling measurement error, informative censoring, and multilevel biomarker structures common in large cohort studies. Further methodological work is also warranted to refine intercept estimation under extreme censoring and to characterize the finite-sample behavior of resampling-based variance estimators. Finally, applying these methods to large-scale omics datasets with pervasive detection limits would provide valuable insight into their performance in complex real-world environments.

\end{document}